\documentclass[prl,reprint, nofootinbib,amsmath,amssymb, aps, floatfix]{revtex4-1}

\usepackage{graphicx}
\usepackage{color}
\usepackage{braket}
\usepackage{cancel}
\usepackage{mathrsfs}
\usepackage{enumitem}
\usepackage{amsthm}
\usepackage{mathtools}

\theoremstyle{remark}

\DeclareMathOperator{\tr }{tr}

\usepackage{bbold}
\usepackage{comment}

\definecolor{darkgreen}{rgb}{0,0.5,0}
\definecolor{darkblue}{rgb}{0,0,0.6}
\definecolor{purple}{rgb}{0.4,.2,0.7}
\usepackage[colorlinks=true,citecolor=darkgreen,linkcolor=black,urlcolor=purple]{hyperref}
\usepackage[export]{adjustbox}

\def\be{\begin{eqnarray}}
\def\ee{\end{eqnarray}}

\newcommand{\lan}{\langle}
\newcommand{\ran}{\rangle}

\newcommand{\bea}{\begin{eqnarray}}
\newcommand{\eea}{\end{eqnarray}}
\def\ben{\begin{equation}}
\def\een{\end{equation}}

     \let\r=v

\def\be{\begin{equation}}
\def\ee{\end{equation}}
\def\ba{\begin{array}}
\def\ea{\end{array}}

\def\ba#1\ea{\begin{align}#1\end{align}}
\def\bs#1\es{\begin{split}#1\end{split}}

\allowdisplaybreaks  

\def \be {\begin{equation}}
\def \ee {\end{equation}}
	
\def \JM#1 {{\color{blue}  JM: #1 }}
\def \AAl#1 {{\color{red}  AA: #1 }}

\begin{document}
\title{Measurements \xcancel{without} \emph{with} probabilities in the final state proposal}
\author{Ahmed Almheiri}
\affiliation{New York University Abu Dhabi, Abu Dhabi , P.O. Box 129188, United Arab Emirates, }
\affiliation{Center for Cosmology and Particle Physics, New York University, New York, NY 10003, USA }

\begin{abstract}

Bousso and Stanford (BS) argued that the black hole final state proposal leads to acausal effects and ill-defined probabilities for the AMPS experiment. We identify a loophole in their analysis using insights from entanglement wedge reconstruction and replica wormholes. We trace the cause of the BS problems to the misidentification of the physical interior where the second AMPS measurement happens from among the multiple interiors introduced by the first measurement.

\end{abstract}
\maketitle

\paragraph{\textbf{Unitary evolution by post-selection}}
The black hole final state proposal \cite{Horowitz:2003he} seemingly evades the information \cite{Hawking:1976ra} and firewall \cite{Almheiri:2012rt} paradoxes  in one fell swoop. But there's a catch: it requires modifying quantum mechanics inside  black holes. The proposal posits a final state boundary condition, usually understood to be imposed at the singularity, that performs a measurement on the state of the interior with a post-selected outcome. Using the entanglement of the Hawking radiation across the horizon as a resource, this measurement quantum teleports \cite{Bennett:1992tv} the information inside the black hole into the outgoing Hawking radiation.

\begin{figure}[t]
 \includegraphics[scale=1.8, valign = c]{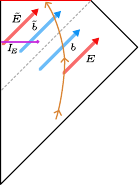}
\caption{\label{fig_amps_experiment}
{\footnotesize An infalling observer (orange) performs experiment {\bf (I)} on the early radiation $E$  and the late mode $b$  before jumping into the black hole and performing experiment {\bf (II)} on the late mode $b$ and it's partner behind the horizon $\tilde{b}$. After the Page time, the island (purple) belongs to the entanglement wedge of the radiation $E$, which may be modified by operations on $E$ such as the first AMPS experiment {\bf (I)}. }}
\end{figure}

The proposal appears to fall short when describing the experience of an infalling observer. Consider the AMPS experiment (see FIG \ref{fig_amps_experiment}) where an infalling observer performs two measurements to verify that: {\bf (I)}  a late Hawking mode $b$ is entangled with the early radiation $E$ as required by unitarity, and {\bf (II)}  the same late mode $b$ is purified by its interior partner $\tilde{b}$ as required by effective field theory in the vicinity of the event horizon  \cite{Almheiri:2012rt}. While such a state of  $\tilde{b}bE$ is precluded by strong subadditivity of entanglement entropy \cite{Almheiri:2012rt,Braunstein:2009my,Mathur:2009hf}, it can be assembled with post-selection \cite{Lloyd:2013bza}. However, Bousso and Stanford (BS) have argued that the extent to which this is true is limited to experiments that exclusively perform {\bf (I)} \emph{or} {\bf (II)} but not both \cite{Bousso:2013uka}.
In the scenario where a single observer does measure both, i.e. one that actually realizes the AMPS thought experiment, BS encounter two difficulties:
\begin{enumerate}[label = {\bf (\Alph*):}]
		\item The results of measurement {\bf (II)} cannot be assigned probabilities.
		\item The statistics of measurement {\bf (I)} depends on whether   {\bf (II)} will be performed in the future.
\end{enumerate}
Difficulty {\bf (A)} occurs when branches of the wavefunction with different measurement outcomes fail to decohere. Working in the decoherence functional formalism \cite{zurek1990complexity,Dowker:1992es,Gell-Mann:1992wkv}, BS find that the decoherence functional of the AMPS measurements, namely the overlap matrix between different possible histories, is not diagonal. The other problem, difficulty {\bf (B)}, is more severe, predicting observable violations of causality outside of black holes. We will refer to {\bf (A)} and {\bf (B)} respectively as the decoherence and acausality problems.

We revisit these problems armed with insights from the Page curve calculation via replica wormholes \cite{Penington:2019npb, Almheiri:2019psf,Penington:2019kki, Almheiri:2019qdq,Almheiri:2019hni,Almheiri:2019psy,Almheiri:2019yqk}(see \cite{Almheiri:2020cfm} for a review). For instance, the sensitivity of the interior to operations on the Hawking radiation, as implied by replica wormholes,\footnote{This property was previously motivated and analyzed in \cite{Verlinde:2012cy,Papadodimas:2012aq,Maldacena:2013xja}} did not factor into the BS analysis. One such operation that features in the first AMPS experiment {\bf (I)} is the distillation of a mode $e_b$ from the early radiation $E$ that is maximally entangled with the late mode $b$. This distillation step was implicit in the BS analysis which, as we will see, introduces a subtlety in how the second AMPS measurement {\bf (II)}  is carried out.

We will analyze this subtlety in a model of the black hole final state that mimics features of replica wormholes while also being explicit with the distillation step and its consequences. We work with a generalization of the random final state model of Lloyd \cite{Lloyd:2004wn} adapted to intermediate times using a non-isometric code model studied in \cite{Akers:2022qdl}. This model is motivated by an observation made by Stanford \cite{Stanford_KITP_Talk} of the similarity between replica wormholes and Weingarten contractions between Haar integrated random unitaries.

In this letter, we perform a detailed analysis of the AMPS experiment within the random final state model. We will show that the first AMPS measurement {\bf (I)} introduces multiple black hole interiors and provide a prescription to select the true interior where the second AMPS measurement {\bf (II)} happens. We will find the results of the AMPS experiment to be consistent with unitarity and the semi-classical description of the horizon. Finally, we turn to the BS implementation of the final state and trace their difficulties to a different and, as we argue, wrong choice of interior for the second measurement.


\paragraph{\textbf{A not so final ``final state''}}
The original black hole final state proposal generates a pure state of the outgoing radiation by projecting the state of everything that will ever fall into the black hole onto a predetermined pure state. To consider experiments on an extant black hole and its radiation, we need a refinement that applies to intermediate times. Such a final state would prepare a pure bipartite state of the outgoing radiation and what can be called the black hole fundamental degrees of freedom.\footnote{As in those that account for the $e^{S_\mathrm{BH}}$ microstates of the black hole.} We introduce such a model in this section. Our focus will be on information-theoretic aspects of the problem and less so on gravitational dynamics, and so our toy model will be described by quantum circuits involving finite dimensional Hilbert spaces. We will comment on  physical implications that go beyond this model as needed.

There are two ingredients going into the black hole final state proposal. The first is the effective field theory description of the state of the quantum fields on a Cauchy slice threading the event horizon of the black hole. Let $| \psi \ran_M$ be the state of a collapsing shell that forms the black hole, and let $|\phi \ran$ be the state of the Hawking radiation. To set up the AMPS experiment, we decompose the Hawking radiation into the early radiation $E$, a single late mode $b$, and their interior partners $\tilde{E},\tilde{b}$ respectively. We will take the state of the Hawking radiation, and its diagrammatic representation, to be
\begin{align}
	|\phi \ran \equiv | \phi_{b} \ran_{\tilde{b}b}| \phi_{E} \ran_{\tilde{E}E} =  \includegraphics[scale=1.1, valign = c]{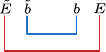},
\end{align}
where the state $| \phi_{A} \ran_{AB}$ is a maximally entangled state of two equal $A$-dimensional Hilbert spaces ${\cal{H}}_A,{\cal{H}}_B$.\footnote{We use the Hilbert space label to refer to its dimension.} Working after the Page time but while the black hole is still macroscopic, we assume that $E\gg M/E \gg b$, where $M$ is the dimension of black hole Hilbert space right after collapse. The property of maximal entanglement here mirrors the maximal entanglement in the thermofield double state at fixed average energy.

The second input into the final state proposal is the final state itself. We consider a partially projected Haar random unitary\footnote{Strict Haar randomness is not needed. It is sufficient to consider a $K$-design ensemble with sufficiently large $K$ that mimics Haar randomness for up to $K$ moments of $U,U^*$.} that takes in the state of everything in the black hole interior and produces a state of the black hole's fundamental degrees of freedom, represented by
\begin{align}
		\lan F|U &\equiv \includegraphics[scale=1, valign = c]{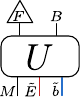}.
		\label{eq_projector}
\end{align}
The bottom legs take in the state of the interior composed of the infalling matter and interior Hawking modes. The upper legs represent the black hole's fundamental degrees of freedom $B$ and the remaining factor projected on by the state $\lan F |$. The details of the projector are not important since they are washed out by the Haar integral over $U$. Its dimension must be $E^2b^2$ to conserve the Hilbert space dimension going from the infalling matter to the Hawking radiation and the remaining black hole. This tensor is an example of a non-isometric code \cite{Akers:2022qdl} that maps  ${\cal H}_M \otimes {\cal H}_{\tilde{E}} \otimes {\cal H}_{\tilde{b}}$ to ${\cal H}_B$ where $B < M E b$.

To obtain the state of the Hawking radiation and the black hole we must apply the map \eqref{eq_projector} onto the state of the infalling matter (initial black hole) and all of the emitted Hawking radiation. Writing the combined state as $|\psi, \phi_{b},\phi_{E} \ran_{M\tilde{b}b\tilde{E}E}$, the projected state is
\begin{align}
	\lan F|U|\psi, \phi_{b},\phi_{E} \ran_{BbE}= \includegraphics[scale=1, valign = c]{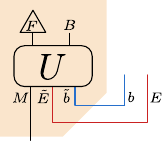}
		\label{fig_projected_state_empty},
\end{align}
where the shaded region (cantaloupe) represents the black hole interior. This prepares a pure state of the tripartite system $BbE$. If we choose to trace out $B$ then one obtains a prescription for  computing the density matrix of the radiation via a {\emph{final density matrix}}.\footnote{We thank Douglas Stanford for discussion on this point.}

\paragraph{\textbf{Distillation}}
For ordinary evaporating black holes past the Page time, the spacetime develops a non-trivial \emph{quantum extremal surface} that extremizes the generalized entropy of the Hawking radiation. This surface bounds a region in the black hole interior, known as the island $I_E$ shown in FIG \ref{fig_amps_experiment}, that contributes to the fine-grained entropy of the radiation. Because of this, the island is said to  belong to the \emph{entanglement wedge} of the Hawking radiation which, by the quantum error correction (QEC) interpretation \cite{Almheiri:2014lwa} of AdS/CFT and entanglement wedge reconstruction \cite{Jafferis:2015del, Dong:2016eik}, should be considered a subsystem of the Hawking radiation. Within this framework, distillation of $e_b$ is simply the extraction of the mode $\tilde{b}$ that purifies $b$ in the semi-classical description.

The black hole final state adopts an equivalent QEC interpretation. The final state projector behaves as an encoding map from the state of the interior to the combined state of the outgoing radiation and the black hole. Operations on $\tilde{b}$, or equivalently $e_b$, can be implemented using a Petz map operator \cite{Petz:1986tvy, Penington:2019kki}. The first step to defining this operator is to define the subspace on which it acts. We consider a \emph{code subspace} spanned by states $|\Psi_i \ran$ where $e_b$ is not distilled and where $i$ labels the state of $\tilde{b}$, and states $|\Psi_0 \ran_{B\hat{E}}|j \ran_{e_b}$ where $e_b$ is distilled and whose states are labelled by $j$.\footnote{The removal of $\tilde{b}$ presumably creates a firewall. However, the details of this will not be important because no observers will be sent into distilled black holes.} Note that  distillation does not change the dimension of the black hole but reduces that of the radiation, and so $\hat{E} = E/b$. The explicit expressions for bras of one and kets of the other are
\begin{gather}
	_{BE}\lan \Psi_i | = \lan \psi, \phi_{E}|\otimes\!  \, _{\tilde{b}}\!\lan i |U^\dagger |F \ran, \\
	|\Psi_0 \ran_{B\hat{E}}|j \ran_{e_b} = \lan F|\widetilde{U}|\psi, \phi_{E/b} \ran \otimes |j \ran_{e_b},
\end{gather}
where the specific $\tilde{U}$ is a choice determined by the decoding operator and  defines the final state of the distilled black hole. Note also that the projector $\lan F |$ here acts on a different dimension Hilbert space than that of $U$; here it acts on a Hilbert space of dimension $E^2 = \hat{E}^2 b^2$ rather than $E^2b^2$.

Working in this code subspace, a candidate distillation operator is simply  $D_\mathrm{code} \equiv \sum_j|\Psi_0 \ran_{{B\hat{E}}}|j \ran_{e_b}\times\, _{BE\! }\lan \Psi_j |$. However, the AMPS experiment requires a distillation operator that acts exclusively on the radiation while $D_\mathrm{code}$ has support on the black hole. The desired operator on the radiation is the Petz-Lite operator \cite{Penington:2019kki} obtained by tracing out the black hole subfactor, $D_E \equiv \tr_B D_\mathrm{code}$, given diagrammatically by
\begin{align}
	D_E = \includegraphics[scale=1.05, valign = c]{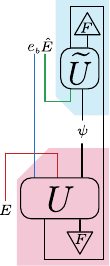}
		\label{eq_decoder}.
\end{align}
This diagrammatic form of the distillation operator makes clear  that multiple black hole interiors are in the game. It also makes clear that the distilled mode $e_b$ is nothing but $\tilde{b}$ from the black hole interior (flamingo) in the bra.

Acting with the distillation operator on the state of the evaporating black hole $\lan F| U | \psi, \phi_b, \phi_E \ran_{BbE}$ generates a spacetime where $b$ is entangled with a mode $e_b$ in the early radiation and removes $\tilde{b}$ from the interior. This is confirmed by the  semi-classical description of the modified state $|{ D}_E\Psi \ran \equiv { D}_E\lan F| U | \psi, \phi_b, \phi_E \ran_{BbE}$ that is assigned by the dominant saddle point of the gravitational path integral computing the norm of the state, $\lan {D}_E\Psi | {D}_E\Psi \ran$.  The semi-classical description is constructed by looking at the geometry on a  time reflection symmetric slice and its semi-classical time evolution. In the model of the final state we are considering, the leading contribution to the overlap, after integrating over the random unitaries, looks like
\begin{align}
	\int \! \!dU\! d\tilde{U} \lan {D}_E\Psi | {D}_E\Psi \ran \approx \includegraphics[scale=1, valign = c]{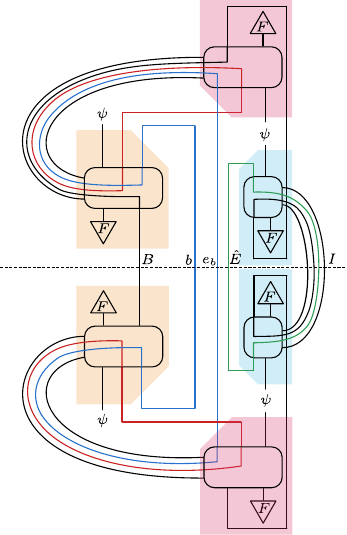}
		\label{eq_distilled_saddle},
\end{align}
where the connection between the unitaries are contractions implemented  by the Haar integrals. This  contribution dominates because it maximizes the number of early radiation loops; unitaries joined by early radiation lines get connected. In a physically realistic model, these connections represent wormholes connecting islands of different black holes \cite{Stanford_KITP_Talk}.

\begin{figure}[t]
	\includegraphics[scale=0.9, valign = c]{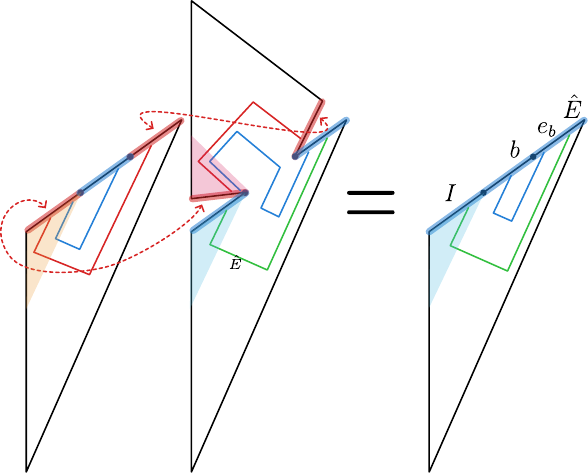}
	\caption{Acting on the original black hole radiation system (left spacetime) with  the distillation operator ${D}_E$ (middle spacetime) produces a new spacetime composed of pieces from the original system and ${D}_E$ (right spacetime). It does so by identifying the red shaded regions of the operator bra and the original state, and stitches the blue regions to produce the state on the right where $b$ is purified by a mode $e_b$ in the radiation.}
		\label{fig_spacetime_dist}
\end{figure}

The spacetime described by the distilled state is constructed from  pieces on the time symmetric slice, as shown by the dashed line in \eqref{eq_distilled_saddle}: it is composed of the original black hole degrees of freedom $B$ and late mode $b$, as well as the distilled mode $e_b$ and the remainder of the Hawking radiation $\hat{E}$. Importantly, the new black hole interior is the one introduced by the distiller, labelled by $I$ in the diagram. A spacetime description of this is presented in Fig \ref{fig_spacetime_dist}.\footnote{The operator $D_E$ is an example of a timefold \cite{Heemskerk:2012mn} contrary to the claim in \cite{Bousso:2025udh} of no such construction for island reconstruction. The simplicity of this construction is due to $D_E$ not being unitary, but a unitary version  of \cite{Yoshida:2017non} also admits a timefold picture but with an exponentially large number of folds.}

\paragraph{\textbf{AMPS experiment with a final state}}
With the distillation operator at hand, we are ready to analyze the AMPS experiment.

In the first AMPS measurement, the infalling observer must distill the mode $e_b$, measure $b e_b$, and then instill (encode) the mode $e_b$ back into $E$. We will compare the probability of $be_b$ being in the state $|\phi \ran_{be_b}$ against all other possibilities. These two options are projected on by $\Pi^{(1)}_{be_b} \equiv \Pi_{be_b} \equiv |\phi \ran \lan \phi |$ and $\Pi^{(2)}_{be_b} \equiv 1 - \Pi_{be_b}$, respectively. To get projectors that act on  $bE$ we must conjugate the $be_b$ projectors with the distiller giving $\Xi^{(1)}_{be_b} \equiv  D_E^\dagger \Pi_{be_b}^{(n)}D_E$ and $\Xi^{(2)}_{be_b} \equiv D_E^\dagger(1- \Pi_{be_b})D_E$.\footnote{ A unitary implementation of the distillation is a computationally complex protocol and takes an exponential-in-system-size amount of time to implement, exponentially larger than is available to the infalling observer \cite{Harlow:2013tf}. Nevertheless, we will ignore this objection and assume that  one of the workarounds proposed in \cite{Oppenheim:2014gfa, Almheiri:2013hfa} is implemented.} Following this measurement, the resulting state $|\Pi^{(n)}\Psi \ran \equiv  \Xi^{(n)}_{bE}\lan F| U | \psi, \phi_b, \phi_E \ran_{BbE}$ has the diagrammatic representation
\begin{align}
	|\Pi^{(n)}\Psi \ran = \includegraphics[scale=1, valign = c]{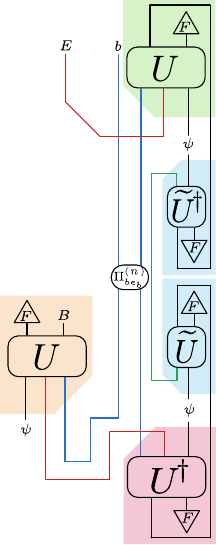}.
		\label{eq_first_amps}
\end{align}

In contrast to the first AMPS measurement, the second  measurement {\bf{(II)}} does not require  distillation; the observer simply falls into the black hole and measures $\tilde{b}b$. But, which black hole? The first measurement introduces four new black hole interiors, and we need a way to determine which of those the observer actually falls into.

We already encountered this problem when analyzing the distiller; the true black hole interior belongs to the spacetime described by the state following the distillation, i.e. the dominant contribution to the norm of the state. Similarly, we are interested  in observations of the black hole in the dominant contribution to the norm of the measured state.\footnote{We thank Xi Dong for discussions on this point} As before, we need to consider the time symmetric slice and ask which black hole interior, or which unitary contraction, does it intersect.

This is easy to determine: in the dominant contribution to the norm, the unitary contractions follow early radiation contractions, and therefore only the last unitary in the first AMPS measurement (sage in \eqref{eq_first_amps}) will intersect the time symmetric slice. Therefore, the physical mode $\tilde{b}$  the observer will encounter is the one that feeds into this unitary. Again, we want the probabilities of $\tilde{b}b$ being in $|\phi \ran_{\tilde{b}b}$ or otherwise, given by the projectors $\Pi^{(1)}_{\tilde{b}b} \equiv \Pi_{\tilde{b}b}$ and $\Pi^{(2)}_{\tilde{b}b} \equiv 1 - \Pi_{\tilde{b}b}$. Then, the resulting state from performing the full AMPS experiment is 
\begin{align}
	|\Pi^{(m,n)} \Psi \ran \equiv \includegraphics[scale=1, valign = c]{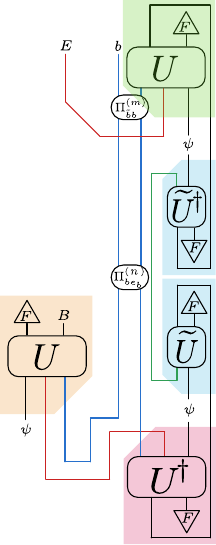}.
		\label{eq_new_amps}
\end{align}
For simplicity, the projection of $\tilde{b}b$ is implemented without dropping $b$ into the black hole. This is a measurement in the semi-classical regime and therefore should be independent of semi-classical evolution through the horizon.\footnote{One could consider a variant of the experiment where $b$ and $\tilde{b} $ are measured in sequence; given many copies, one checks that measurements of $b$ and $\tilde{b}$ agree in different bases.} States with different values of $m,n$ represent different histories, and the (normalized) decoherence functional is simply the overlap matrix ${\cal D}_{(m,n)}^{(m', n')} \equiv \lan\Pi^{(m',n')} \Psi |\Pi^{(m,n)} \Psi \ran/\lan \Psi | \Psi \ran$.


It is  very straightforward to compute the Haar average of the (normalized) decoherence functional  ${\cal D}_{(m,n)}^{(m', n')}$. Since the two measurements occur sequentially on the same systems, their results must match, and so $|\Pi^{(m,n)} \Psi \ran = \delta_{mn}|\Pi^{(n,n)} \Psi \ran$.  Moreover, since the distiller successfully distills, we have  $\delta_{mn}|\Pi^{(n,n)} \Psi \ran \approx \delta_{mn}\delta_{n1}|\Pi^{(1,1)} \Psi \ran \approx \delta_{mn}\delta_{n1}| \Psi \ran$ to exponential accuracy (averaging over the final state is implicit). Therefore, the only nonzero component of the decoherence functional is ${\cal D}_{(1,1)}^{(1, 1)}$ and so is diagonal by default. It is also equal to 1, indicating both AMPS measurements succeed with 100\% certainty.

Therefore we find that the AMPS experiment confirms unitarity and smoothness of the horizon without leading to any acausality or ill-defined probabilities.

\paragraph{\textbf{AMPS experiment, the BS version}}
We now debug the BS version of the AMPS experiment. Their prescription can be framed entirely within the above model  but with drastically different, and troublesome, conclusions. The key difference from our analysis will be the choice of black hole interior  probed by the second AMPS experiment. We argue that the BS choice is not consistent with lessons from replica wormholes and entanglement wedge reconstruction  and, to some extent, quantum mechanics itself.

The flaw is a too literal interpretation of the original black hole final state proposal. The original implementation begins with the semi-classical state of the collapsing matter and Hawking radiation,  projects onto a given history, and then follows with application of which the final state. For the AMPS experiment, this reads
\begin{align}
	|\Pi^{(m,n)} \Psi \ran \equiv  \lan F| U \, \Pi_{\tilde{b}b}^{(m)}\Xi^{(n)}_{bE}| \psi,\phi_{b}, \phi_{E}\ran.
	\label{BSstate}
\end{align}
This is the BS version of the AMPS experiment in the language of our model. This prescription is agnostic to the details of the projectors involved. 

There is something immediately suspect about the state \eqref{BSstate}. In particular, it privileges the original final state over those appearing in the first AMPS measurement projector $\Xi^{(n)}_{bE}$. This would be fine if the history involved only semi-classical measurements, which do not involve extra copies of final state, where the original final state is privileged by default. This can be confirmed using  methods of the previous sections;  semi-classical projectors do not change the semi-classical saddle, and the original interior remains the true interior.

The original prescription enforces the mode $\tilde{b}$ of the second AMPS experiment to be the one inside the original black hole interior;  the diagrammatic representation of the measured state is
\begin{align}
	|\Pi^{(m,n)} \Psi \ran =\includegraphics[scale=0.95, valign = c]{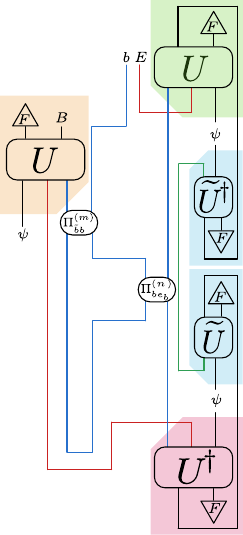}
		\label{eq_BS_amps}.
\end{align}
where the second AMPS experiment projector acts on the $\tilde{b}$ system inside the original interior (cantaloupe). Therefore, the observer in this version of the AMPS experiment falls into the original black hole interior despite the action of the Petz operator.

This prescription empties entanglement wedge reconstruction of any meaning. Petz type operators modify the state of the entanglement wedge by projecting out the state describing the original spacetime and replaces it with a new state describing a modified spacetime. Following such an operation, the BS prescription would imply that further probes see only the original spacetime and not the new one. In fact, this is not consistent with how quantum mechanics works; operators change states by projecting them out and introducing new ones, and this prescription would render such changes unobservable.

It also explains the acausality encountered by BS; the second measurement probes the state prior to the action of the first. This can be seen in the decoherence functional whose leading contribution after performing the Haar integral is
 \begin{gather}
 	\int\: \! \!\! dU d\tilde{U} \, {\cal D}_{(m,n)}^{(m', n')} \approx \includegraphics[scale=0.8, valign = c]{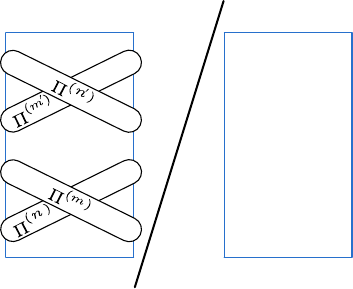}.
 \end{gather}
 This shows that the result can be expressed as the decoherence functional of a history of measurements on two systems with no obvious time ordering; the measurements are crisscrossed.  Evaluating this reproduces the decoherence functional of BS \cite{Bousso:2013uka}.

\paragraph{\textbf{Final words}}

What makes the problem of the black hole interior so difficult is the lack of a non-perturbative definition of observables behind the horizon. Observables in quantum gravity are defined relative to some fixed structure, usually at infinity, and therefore are never guaranteed to probe the interior of black holes.\footnote{See \cite{AliAhmad:2025kki} on surmounting this by placing the holographic boundary behind the horizon using a generalization of the $T\bar{T}$ deformation \cite{Smirnov:2016lqw, Cavaglia:2016oda}.} It is only in the semi-classical description that the notion of the interior is sharp.

This makes the black hole final state proposal  worth fighting for. It retains the semi-classical picture while also incorporating quantum gravitational effects using the final state. It gives hope that quantum gravity may be formulated using familiar notions like geometry, spacetime, and horizons.

Our results support the final state proposal by showing that the AMPS experiment produces results consistent with all postulates of complementarity. The key was to provide a self consistent implementation of the final state that works for operations that go beyond the semi-classical description, as is involved in the first AMPS measurement. Furthermore, we successfully debugged the Bousso and Stanford analysis, showing that their implementation of the final state misidentifies where the second AMPS measurement happens.

\paragraph{\textbf{Acknowledgements}}
I am indebted to Raphael Bousso, Ven Chandrasekaran, Xi Dong, Adam Levine, Douglas Stanford, Misha Usatyuk, Ying Zhao for extensive discussions. I also thank Douglas Stanford and Ven Chandrasekaran for separate collaborations on aspects of this project. I thank Shadi Ali Ahmad, Simon Lin and Shoy Ouseph for reading a version of this draft. This work was supported by Tamkeen under the NYU Abu Dhabi Research Award (ADHPG
-- AD375).

\bibliographystyle{ourbst}
\bibliography{post.bib}


\end{document}